\documentclass[11pt,prd,showpacs,showkeys]{revtex4}
\usepackage{amsmath}
\usepackage{epsfig}
\usepackage{graphicx}
\usepackage{color}
\begin{document}
\title{
Neutrino masses and tribimaximal mixing in
the minimal renormalizable
\\SUSY $SU(5)$ Grand Unified Model
with  $A_4$ Flavor symmetry}
\author{Paolo Ciafaloni}
\author{Marco Picariello}
\author{Emilio Torrente-Lujan}
\author{Alfredo Urbano}
\email{Paolo.Ciafaloni@le.infn.it, Marco.Picariello@le.infn.it, etl@um.es, Alfredo.Urbano@le.infn.it}
\affiliation{\ }
\affiliation{Dipartimento di Fisica - Universit\`a del Salento, Lecce, Italia}
\affiliation{Istituto Nazionale di Fisica Nucleare - Lecce, Italia\\}%
\affiliation{Dep. de Fisica, Grupo de Fisica Teorica, Univ. de Murcia - Murcia, Spain.}
%EndAName

\begin{abstract}
We analyze all possible extensions of the recently proposed
minimal renormalizable SUSY $SU(5)$ grand unified model
with the inclusion of an additional $A_4$ flavor symmetry.
We find that there are 5 possible Cases but only one of them is
phenomenologically interesting. We develop in detail such Case and
we show how the fermion masses and mixing angles come out.
As prediction we obtain the neutrino masses of order of $0.1\,eV$
with an inverted hierarchy.
\end{abstract}
\pacs{11.30.Hv, 12.10.-g, 14.60.Pq, 12.15.Ff}
\keywords{Flavor symmetries, Unified field theories and models,
Neutrino mass and mixing, Quark and lepton masses and mixing}
\maketitle

\newcommand\B[1]{\overline{#1}}

%%%%%%%%%%%%%%%%%%%%%%%%%%%%%%%%%%%%%%%%%%%%%%%%%%%%%%%%%%%%%%%%%%%%%%%%
\section{Introduction}
%%%%%%%%%%%%%%%%%%%%%%%%%%%%%%%%%%%%%%%%%%%%%%%%%%%%%%%%%%%%%%%%%%%%%%%%%%
The flavor puzzle is one of the most intriguing problem in
particle physics. Questions like  whether  there is any reason behind the charged
fermion  mass hierarchy, why the quark mixings
are so small while two of the lepton mixings are large or
 why one lepton mixing is so close to
be maximal while the other large angle is not are still far to be satisfactorily solved.

These problems have usually been approached all together
at the same time and  a single mechanisms have been suggested for their solution.
Recently
\cite{Bazzocchi:2008rz}-\cite{Bazzocchi:2008ej}
%\cite{Adulpravitchai:2008yp,Seidl:2008yf,Hagedorn:2008bc,Araki:2008rn,Kobayashi:2008ih,Feruglio:2008py,Feruglio:2008ht,Mitra:2008bn,Lin:2008aj,Adhikary:2008au,Plentinger:2008nv,Araki:2008xq,Bazzocchi:2008sp,Lam:2008sh,Escobar:2008vc,Luhn:2008sa,Haba:2008dp,Ding:2008rj,Bazzocchi:2008rz,HernandezGaleana:2007xs,Rodejohann:2008xp,Rodejohann:2007zz,Bazzocchi:2007au,Hochmuth:2007wq,Picariello:2007ss,Law:2009vh,Chen:2008eq,Everett:2008et,Hirsch:2008mg,Albright:2008qb,Parida:2008pu,Hirsch:2008rp,Azatov:2008vu,Plentinger:2008up,Stech:2008wd,Altarelli:2008bg,Raidal:2008jk,Altarelli:2007gb,Pakvasa:2007zj,Chen:2007gp,Grimus:2008tm,Luhn:2007yr,Bazzocchi:2007na,Plentinger:2007px,Luhn:2007sy,Chen:2007jsa,Altarelli:2007cd,Chen:2007afa,Ma:2007ia,Picariello:2007yn,McKeen:2007ry,Feruglio:2007uu,Rodejohann:2007zza,Chen:2007zj,Agarwalla:2006dj,Duret:2008st,Niehage:2008sg,Nandi:2007cw,NimaiSingh:2007zb,Duret:2007ux,Picariello:2006sp,Schmidt:2006rb,Hochmuth:2006xn,Strumia:2006db,Ishimori:2008fi,Ishimori:2008uc,Bazzocchi:2008ej,Feruglio:2007hi,Morisi:2007ft,Caravaglios:2005gf,Altarelli:2005tj,Altarelli:2005yp,Fogli:2008cx,Das:2000uk,Slansky:1981yr,Chauhan:2006im}
it has become manifest that the three flavor problems have to be approached
in a different way and must be solved by introducing
different mechanisms \cite{Picariello:2006sp,Picariello:2007yn} for each of them.
In particular, while hierarchies are described by continuous symmetries,
the mixings can be explained by introducing discrete symmetries
\cite{Bazzocchi:2007au,Bazzocchi:2008rz}.
%%%%%%%%%%%%%%%%%%%%%%%%%%%%%%%%%%%%%%%%%%%%%%%%%%%%%%%%
% S(3) x P discrete symmetry
For example, in
\cite{Picariello:2006sp}-\cite{Mitra:2008bn}
%\cite{Altarelli:2005yp,Altarelli:2005tj,Caravaglios:2005gf,Picariello:2006sp,Morisi:2007ft,Picariello:2007yn,Ma:2007ia,Picariello:2007ss,Altarelli:2007gb,Feruglio:2007hi,Raidal:2008jk,Altarelli:2008bg,Mitra:2008bn},
several attempts have been done to face the flavor puzzle by introducing
flavor symmetries based in discrete finite groups
 such as $S_3$, $S_4$, $A_4$, $T'$, and so on.

%%%%%%%%%%%%%%%%%%%%%%%%%%%%%%%%%%%%%%%%%%%%%%%%%%%%
% A(4) into left-right flavor symmetry
Moreover some attempts
\cite{Bazzocchi:2008rz,Luhn:2008sa,Lin:2008aj,Haba:2008dp,Adhikary:2008au}
have been done to embed the discrete symmetery, as the $A_4$ one,
into larger, continuos groups to explain also the hierarchy
among the 3rd and the other two generations.
It is has been shown there how  the discrete symmetry $A_4$ can help  in
solving both aspects of the flavor problem:
lepton-quark mixing hierarchy and family mass hierarchy.
%%%%%%%%%%%%%%%%%%%%%%%%%%%%%%%%%%%%%%%%%%%%%%%%%%%
% A(4) into SU(3) x U(1) flavor symmetry
The  flavor symmetry $A_4$, as shown for example in
\cite{Bazzocchi:2008rz,Plentinger:2008nv,Adhikary:2008au,Lin:2008aj,Mitra:2008bn},
is also a very promising option if extended
to a larger flavor group compatible with
$SO(10)$-like gauge grand unification. For example, by embedding
$A_4$ into a group like $SU(3)\times U(1)$ \cite{Bazzocchi:2008rz},
it is possible to explain both large neutrino mixing and fermion
mass hierarchy in a $SO(10)$ Grand Unified Theory of Flavor (GUTF).
%%%%%%%%%%%%%%%%%%%%%%%%%%%%%%%%%%%%%%%%%%%%%%%%%%%

The embedding of discrete symmetries in a grand unified theories is
somehow simpler in $SU(5)$
GUTS \cite{Chen:2007afa,Chen:2007gp,Altarelli:2008bg,Plentinger:2008nv}.
In fact in this case matter fields belong to
two different representations ${\bf 10_T}$ and ${\bf \B5_T}$ under the
gauge group. So there is one degree of freedom more available in the flavor transformations.

Recently a supersymmetric renormalizable grand unified theory
has been proposed in \cite{Perez:2007iw} based on the $SU(5)$ gauge
symmetry where the neutrino masses are generated through
type I and type III seesaw mechanisms.
Within this model it is possible to generate all fermion masses
with the requested minimal Higgs sector,  the existence of one
massless neutrino is predicted and the leptogenesis mechanism can successfully be realized.
Moreover it has been shown that  the predicted decay of the proton and neutralino properties
 are in agreement with  experimental evidence.
This theory can be considered  a simple renormalizable
supersymmetric grand unified theory based on the $SU(5)$ gauge
symmetry since it has the minimal number of superfields and
free parameters.

In the framework of SUSY $SU(5)$ GUT models, we  will investigate in this work
the possibility of adding an $A_4$ flavor symmetry
   to constrain the flavor structure of
the coupling constants. We will verify  the predictions
on the quark and lepton mixing angles obtained from it as well as the possibility
to accommodate all the fermion masses and mixing.

We start considering a class of models requested to satisfy the
following conditions:
\begin{itemize}
\item minimal, in the sense that they contain the minimal number
 of superfields (without any gauge singlet) compatible with experimental evidence;
\item renormalizable, in the sense that we include all the operators
 of dimension less or equal to four compatible with our symmetry;
\item supersymmetric, where SUSY is helpful in fixing  the vacuum alignments;
\item Grand Unified, where the  gauge group is
 $SU(5)$, and flavor symmetric, where the flavor group is $A_4$.
\end{itemize}
Remarkably, there are only a few  ``Cases'', according to the field transformations,
which are compatible with these requirements.
Realistic models are selected by imposing that
the actual fermion masses can be reproduced, i.e. there are not
  charged fermions with zero mass and  at least two neutrinos
  are massive;
the CKM mixing matrix is not the identity at tree level;
 the PMNS mixing matrix is close to the tribimaximal one.
We will  end up with the non trivial result that there is only one
possible model or ``Case'' which  is compatible with the experimental evidence.

In the next section we will present the field content of the model and detail the
transformation properties of the fields according to the gauge and flavor groups for
each of the initial five cases.

\section{Field content and $SU(5)\times A_4$ invariance}

We start considering the field content of a simple
 renormalizable supersymmetric
grand unified theory based on the $SU(5)$ gauge symmetry (the supersymmetric adjoint $SU(5)$ proposed in \cite{Perez:2007iw}). This comprises
the following chiral superfields:
$%$
{\bf10_T},\, {\bf\B5_T},\, {\bf24_T},\,
{\bf\B5_H},\, {\bf\B{45}_H},\, {\bf 5_H},\, {\bf 45_H},\, {\bf 24_H}\
$%$
where the subindex $_{\bf{T}}$ refers to matter fields, defined   respect to
R-parity, and $_{\bf{H}}$ to Higgs; the index number is the representation
dimension under $SU(5)$.
The renormalizable operators allowed under  $SU(5)$
invariance appearing in the supersymmetric
superpotential are \cite{Perez:2007iw}:
%%%%%%%%%%%%%%%%%%%%
\begin{subequations}\label{eq:operators}
%%%%%%%%%%%%%%%%
\begin{eqnarray}
W_0&=&y_1\, {\bf10_T}\,{\bf\B5_T}\,{\bf\B5_H}
    + y_2\, {\bf10_T}\,{\bf\B5_T}\,{\bf\B{45}_H}
    + y_3\, {\bf10_T}\,{\bf10_T}\,{\bf5_H}
    + y_4\, {\bf10_T}\,{\bf10_T}\,{\bf45_H},\\
W_1&=&\gamma\, {\bf\B5_T}\,{\bf24_T}\,{\bf45_H}
    + \beta\, {\bf\B5_T}\,{\bf24_T}\,{\bf5_H},\\
W_2&=&m_\Sigma\, {\bf24_H}\,{\bf24_H}
    + \lambda_\Sigma\, {\bf24_H}\,{\bf24_H}\,{\bf24_H}
    + m\, {\bf24_T}\,{\bf24_T}
    + \lambda\, {\bf24_T} {\bf24_T} {\bf24_H},\\
W_3&=& m_5\,{\bf\B5_H} {\bf5_H}
    + \lambda_H\, {\bf\B5_H} {\bf24_H} {\bf5_H}
    + c_H\, {\bf\B5_H} {\bf24_H} {\bf45_H} +
\nonumber\\&&\quad\quad
    + b_H\, {\bf\B{45}_H}\,{\bf24_H}\,{\bf5_H}
    + m_{45}\, {\bf\B{45}_H}\,{\bf45_H}
    + a_H\, {\bf\B{45}_H}\,{\bf45_H}\,{\bf24_H}\,,
\end{eqnarray}
%%%%%%%%%%%%%%%%
\end{subequations}
%%%%%%%%%%%%%%%%%%%%
where $\gamma$, $\beta$, $a_H$, $b_H,c_H$
and the $y'$s, $\lambda'$s and $m'$s are  coupling
 constants.
The usual decomposition of the fields under the Standard Model
gauge group is reported in \cite{Slansky:1981yr}.
Let us go now to study the transformation properties with respect the
flavor discrete symmetry.
We need to fix the  $A_4$ representations assigned to the fields and
the $A_4$ directions of the vevs of the Higgs scalars.
For the sake of minimality, we impose these simple assumptions:
a) the fields are either singlet or triplets of $A_4$,
%%%%%%%%%%%%%%%%%%%%%%%%%%%%%%%%%%%%%
b)  ${\bf 24}_H$ is flavor singlet,
%%%%%%%%%%%%%%%%%%%%%%%%%%%%%%%%%%%%%
and
c) all the operators in eqs. (\ref{eq:operators}) are allowed by the
flavor symmetry.
Under these,  there are only 5
possibilities and they are listed in table {\ref{tab:A4all}}.

The trivial Case 1, where $A_4$ does not play any role,
is the one discussed in \cite{Perez:2007iw} and in this case
there is no symmetry in the flavor structure
of the mass matrices of the fermions.
In all the other cases ${\bf10_T}$ must be a triplet under $A_4$.
Then there is a  case where there are three ${\bf\B5_T}$  singlets (possibly
corresponding to different $1$ representations)
and all the other fields, with the exception of the ${\bf24_H}$
that is assumed to be singlet always, are triplets (Case 2);
the case where ${\bf24_T}$ is a singlet and
all the other fields are triplets (Case 3);
the case where all the Higgs fields are singlets
while the matter fields are triplets (Case 4);
finally we have the case where both
matter and Higgs fields are triplet (Case 5).
%%%%%%%%%%%%%%%%%%%%%%%%%%%%%%%%%%%%%%%%%%%%%%%%%%%%%%%%%%%%%%%%%%%%%%%%%%
\begin{table}[hbt]
\begin{center}
\begin{tabular}{||l||c|c|c|c|c|c|c|c||}
\hline\hline
   &${\bf10}_{\bf T}$ & ${\B{\bf5}}_{\bf T}$ & ${\bf24}_{\bf T}$ & ${\B{\bf5}}_{\bf H}$ & ${\B{\bf45}}_{\bf H}$ & $\bf5_H$ & ${\bf{45}}_{\bf H}$ & ${\bf24}_{\bf H}$\\
\hline
\hline
Case 1&1&1&1&1&1&1&1&1\\
\hline
Case 2&3&1&3&3&3&3&3&1\\
\hline
%Case 2b&3&1&3&3&3&3&3&3\\
%\hline
Case 3&3&3&1&3&3&3&3&1\\
\hline
Case 4&3&3&3&1&1&1&1&1\\
\hline
Case 5&3&3&3&3&3&3&3&1\\
%\hline
%Case 3b&3&3&3&3&3&3&3&3\\
\hline\hline
\end{tabular}
\caption{$A_4$ transformation assignments for matter and
Higgs fields in the cases where all the operators in eq.
(\ref{eq:operators}) are allowed.
Moreover we assume minimality in the Higgs sector,
i.e. if a given Higgs transforms as $1$ than there is
only one of it.
For the fields ${\bf 10_T}$ and ${\bf \B 5_T}$ ``$1$''
possibly means three fields, each one transforming as the inequivalent
$1$,$1'$, or $1''$ representations.
}
\label{tab:A4all}
\end{center}
\end{table}
%%%%%%%%%%%%%%%%%%%%%%%%%%%%%%%%%%%%%%%%%%%%%%%%%%%%%%%%
An extra freedom in the definition of the models corresponds to
 the $A_4$-direction of the vevs of the Higgs scalars.
It is well known \cite{Altarelli:2007gb} that
the $A_4$ symmetry in  a triplet representation can get a vev either
by respecting $Z_3$ or $Z_2$,
%only with vevs in the direction
%invariant under the $A_4$  operators $T$ or $S$,
i.e. in the $(1,1,1)$ direction in the first case or in a direction
where only one component is not zero like $(1,0,0)$ in the second case.
As we will show, the only phenomenologically viable
scenario is when both the $\bf{45_H}$ and $\bf{5_H}$ fields are $A_4$-triplets
and their vevs are in two different, $T$ or $S$ compatible, directions,
i.e. ${\bf \hat n_{111}}=(1,1,1)$  and ${\bf \hat n_{1}}\simeq (1,0,0)$
respectively.
The full study of the problem of vacuum aligment in the Higgs sector is beyond the scope
of this work and  will be investigated in detail in a forthcoming paper \cite{next}
 by minimizing the Higgs potential.

Our objective in this work is mainly  to check all different
$A_4$ assignment for  the fields (Cases 1-5 of the table) and study
which of then can be phenomenologically interesting.
In sec. {\bf\ref{sec:Pred}} we will investigate the predictivity of the
model and we show that, despite the higher number of fields introduced,
the flavor symmetry strongly constraints the physical observables
even in the broken phase.
In sec. {\bf\ref{sec:charged}}
we will first investigate the general structure of the mass matrices coming from the
flavor symmetry invariance properties of the mass operators and then
we investigate the   fermion mass and mixing
structure for each case.
In section ${\bf IV}$  we consider the flavor predictions
for the quark mixing and charged leptons while
section  ${\bf V}$  is specifically dedicated to the neutrinos and the full leptonic mixing matrix.
Our results concerning the viability of the different cases
can be summarized as follows:
%%%%%%%%%%%%%%%%%%%
\begin{description}
\item[Case 1.]{In this trivial case $A_4$ does not play any role and
the flavor structure is not constrained by the flavor symmetry.}
\item[Case 2.]{Depending on the transformation of the three ${\bf\B5_T}$ fields, we have different situations.
If they all transform in the same way, i.e. as $1$, then the down sector
mass matrix $M_D$ and charged lepton mass matrix $M_E$ have one zero
eigenvalue, for any vev direction.
If two or three {$\bf\B5_T$} transform
differently each other (i.e. as $1, 1', 1''$), then $M_D$ and $M_E$ have three non zero
eigenvalues and we get in principle a phenomenologically interesting case.}
\item[Case 3.]{The $M_D$, $M_E$ and $M_U$ matrices
have three independent eigenvalues if
and only if one vev is in the direction $(1,1,1)$ and another vev is in the direction
$(1,0,0)$, however the predicted lepton mixing matrix is not
of  phenomenological interest.}
\item[Case 4.]{The CKM mixing matrix is diagonal, the neutrino mass matrix, $M_\nu$, has two zero eigenvalues.}
\item[Case 5.]{
The three up sector quark masses can be reproduced only
if the {$\bf{45}_{\bf{H}}$} and {$\bf{5}_{\bf{H}}$} acquire vev in the direction
$(1,0,0)$ apart from a small correction for the ${\bf{45}_{\bf{H}}}$ vev
like $(1,\epsilon,\epsilon)$.
$M_\nu$ has at least two non zero eigenvalues
while $M_D$, $M_E$ have three independent eigenvalues.
The CKM matrix contains the Cabibbo angle and the lepton mixing matrix
is almost tribimaximal.
}
\end{description}
%%%%%%%%%%%%%%%%%%%%
We conclude that only Cases 2 and 5 are of phenomenological interest.
Of these two the mass matrices for the Case 2, where the
three ${\bf \B 5_T}$ fields transform
differently, have been already extensively discussed in literature
\cite{Mitra:2008bn,Luhn:2008sa,Albright:2008qb,Parida:2008pu,Lin:2008aj,Adhikary:2008au,Plentinger:2008nv,Ding:2008rj,Plentinger:2008up,Bazzocchi:2008rz} in similar, although not
identical settings.
So the only case which present promising novelties and should
 further investigated is Case 5.

\section{Flavor structures from symmetry: masses and mixings}\label{sec:Masses}
In this section we will first concentrate our attention
on the charged fermion mass matrices. As we will see
in sec. {\bf\ref{sec:charged}}, there are three possible
flavor structures for the charged fermion masses, depending
on the  transformation properties under $A_4$ of the Higgs fields.

Then we will investigate the resulting phenomenology in each of the Cases listed
in table (\ref{tab:A4all}). Any model should at least accomplish the
following conditions to be phenomenologically relevant:
a) the charged fermion masses are not degenerated,
b) at least two neutrino masses are not zero,
c) the lepton mixing is almost tribimaximal and
the quark mixing matrix is not diagonal and compatible with the Cabibbo angle.
These statements strongly constrain the $A_4$ assignments.

First we will eliminate Case 4 by using the Cabibbo angle constraint.
Then we will investigate Cases 2 and 3: by asking non zero charge fermion
masses, we obtain that the vev of the $A_4$ triplets must be in two
different directions.
In the Case 2 it is not possible to obtain the non
degenerate spectrum of the leptons and quarks
with the exception when the three matter fields $\bf\B5_T$
transform as $1$, $1'$, and $1''$ under the $A_4$ flavor symmetry.
In the Case 3 we obtain a reasonable fermion spectrum,
 a good quark mixing matrix, but a wrong lepton mixing matrix.

Finally we will study the remaining Case 5 and we will show that, to have two non
zero neutrino masses, the vev of the $A_4$ triplets must be in two
different directions. This constraint automatically guarantees non degenerated
charged fermion masses, quark mixing compatible with the Cabibbo structure,
and a lepton mixing matrix almost tribimaximal.

\subsection{Predictivity and degrees-of-freedom counting}\label{sec:Pred}
The coupling constants with three $A_4$ indices,
i.e. $\gamma^{abc}$, $\beta^{abc}$, $\lambda_\Sigma^{abc}$,
$\lambda^{abc}$, $c_H^{abc}$, and $a_H^{abc}$, have 27 ($=3^3$)
elements each. However, as shown in the Appendix of
ref. \cite{Morisi:2007ft}, eq. (A2),
any of the coupling constants with three $A_4$ indices contains only two
independent elements. For example, in the so-called S-diagonal base, we have
%%%%%%%%%%%%%%%%
\begin{eqnarray}\label{eq:contractions}
\gamma^{abc}\ {\bf\B5_T}^a\ {\bf24_T}^b\ {\bf45_H}^c &=&
  \gamma_1\left(
              {\bf\B5_T}^2\ {\bf24_T}^3\ {\bf45_H}^1
            + {\bf\B5_T}^3\ {\bf24_T}^1\ {\bf45_H}^2
            + {\bf\B5_T}^1\ {\bf24_T}^2\ {\bf45_H}^3\right)
\nonumber\\&&\quad
+ \gamma_2\left(
              {\bf\B5_T}^3\ {\bf24_T}^2\ {\bf45_H}^1
            + {\bf\B5_T}^1\ {\bf24_T}^3\ {\bf45_H}^2
            + {\bf\B5_T}^2\ {\bf24_T}^1\ {\bf45_H}^3\right)\,,
\end{eqnarray}
%%%%%%%%%%%%%%%%
and similarly for $\beta^{abc}$, $\lambda_\Sigma^{abc}$,
$\lambda^{abc}$, $c_H^{abc}$, and $a_H^{abc}$. The 27 coupling constants have been
reduced to only two corresponding to odd and even  $S_3$ permutations.

One could worry about the fact that we are
introducing in our model extra Higgs and fermion fields ${\bf 24_T}$
with respect to the minimal SUSY $SU(5)$.
Notice however that these extra fields do not reduce
the model predictivity  because of the inclusion of the $A_4$ symmetry.
In fact %the model in \cite{Perez:2007iw}
the "Supersymmetric Adjoint $SU(5)$" contains
four 3-by-3 matrices, two 3-vectors, and nine other constants,
with a total of 51 parameters (most of them cannot be observed
at low energy).
Any of the models considered here  contains
less parameters, as a consequence of the
general feature of the flavor symmetries.
For example, as we have seen above, the 3-by-3 matrix of \cite{Perez:2007iw}
for the operator in eq. (\ref{eq:contractions})
translate into the 2 coefficients $\gamma_1$ and $\gamma_2$.
In the same way, it should be observed that the number of Higgs vevs
in our model is the same as the number in  corresponding model without
flavor symmetry \cite{Perez:2007iw}.
This is because $A_4$ triplets can get a vev either
by respecting $Z_3$ or $Z_2$, i.e. each vev
will only depend on one parameter.

\subsection{Charged fermion mass matrices}\label{sec:charged}
We investigate in this section the general flavor structure of the coupling
constants resulting in the mass matrices for the charged matter fermions
after symmetry breaking.

It is instructive to see how we can extract important information about
the mass matrices simply by examining
all the possible situations of Higgs-matter coupling with respect to
the $A_4$ assignments and how is the structure of
Higgs vacuum alignments. The case 5 will be investigated in further
detail and finally its explicit mass matrices will be obtained.

\subsubsection{Mass matrices general structures from $A_4$}
After symmetry breaking, we have three possible type of flavor structures
of the coupling constants, depending on  the $A_4$  transformation properties of the initial
cubic Higgs-Matter (TTH) operators. It is instructive to list all the cases:
%%%%%%%%%%%%%%%%%%%
\begin{description}
\item[Type I. The Higgs is an $A_4$-singlet.] Here,
 the mass matrix results from the   coefficients
of the product of a $A_4$ singlet (the Higgs) with two $A_4$ triplets
(matter fields).
In the Case 4, for example, the following operators appear:
\begin{eqnarray}
      y_1\ {\bf10_T}^a\ {\bf\B5_T}^a\ {\bf\B5_H}\,,
&\quad&
      y_2\ {\bf10_T}^a\ {\bf\B5_T}^a\ {\bf\B{45}_H}\,,
\quad
      y_3\ {\bf10_T}^a\ {\bf10_T}^a\ {\bf5_H}\,,
\quad
      y_4\ {\bf10_T}^a\ {\bf10_T}^a\ {\bf45_H}\,.
\end{eqnarray}
In this situation, the resulting mass matrices are diagonal with elements
proportional to $1$, $\omega$, and $\omega^2$ (where $\omega^3=1$) depending
on the singlet $A_4$ properties.
%%%%%%%%%%%%%%%%%%%%%%%%%%%%%%%%%%%%%%%%%%%%%%%%%%%%%%%%%%%%%%%%%%%%%%%%%%%%
\item[Type II. The Higgs is a triplet and one fermion field contains three $A_4$-singlets.] The mass matrix comes from the coefficients
of the product of two triplets (one Higgs and one matter field) with a set of
singlets (the other matter fields).
This situation appears in Case 2, where we have operators as:
\begin{eqnarray}
     y_1^{i}\ {\bf10_T}^a\ {\bf\B5_T}^i\ {\bf\B5_H}^a\,,
&\quad\quad&
     y_2^{i}\ {\bf10_T}^a\ {\bf\B5_T}^i\ {\bf\B{45}_H}^a\,.
\end{eqnarray}
We can distinguish three subcases for the resulting mass matrix.

In the first subcase (IIA), under the hypothesis that the three singlets
transform in the same way, the corresponding contribution to
charged fermion mass matrix is of the form:
%%%%%%%%%%%%%%%%%%%%%%%%%%%%%%%%%%%%%
\begin{subequations}\label{eq:OneIndex}
\begin{eqnarray}\label{eq:OneIndexOne}
M_f & \sim &
\begin{pmatrix}
v_1\,y^{1}&v_1\,y^{2}&v_1\,y^{3}\cr
v_2\,y^{1}&v_2\,y^{2}&v_2\,y^{3}\cr
v_3\,y^{1}&v_3\,y^{2}&v_3\,y^{3}
\end{pmatrix},
\end{eqnarray}
%%%%%%%%%%%%%%%%%%%%%%%%%%%%%%%%%%%%%
where $v_1$, $v_2$, and $v_3$ are the vevs of the Higgs $A_4$-triplet, while the $y^{i}$ are the three coupling constants.
Notice that the charged lepton and quark mass matrices are  transpose of each other, due to the generic $SU(5)$ properties.

In the second subcase (IIB), supposing that only two of the three
singlets transform in the same way, we have a contribution of the form:
%%%%%%%%%%%%%%%%%%%%%%%%%%%%%%%%%%%%%
\begin{eqnarray}\label{eq:OneIndexTwo}
M_f & \sim &
\begin{pmatrix}
v_1\,y^{1}&v_1\,y^{2}&           v_1\,y^{3}\cr
v_2\,y^{1}&v_2\,y^{2}&\omega^2\, v_2\,y^{3}\cr
v_3\,y^{1}&v_3\,y^{2}&\omega\,   v_3\,y^{3}
\end{pmatrix}\,.
\end{eqnarray}
%%%%%%%%%%%%%%%%%%%%%%%%%%%%%%%%%%%%%

Finally, in the third subcase (IIC), supposing that
three singlets transform in different ways,
i.e. as 1, 1' and 1'', the mass matrix is of the form:
%%%%%%%%%%%%%%%%%%%%%%%%%%%%%%%%%%%%%
\begin{eqnarray}\label{eq:OneIndexThree}
M_f & \sim &
\begin{pmatrix}
v_1\,y^{1}&           v_1\,y^{2}&           v_1\,y^{3}\cr
v_2\,y^{1}&\omega\,   v_2\,y^{2}&\omega^2\, v_2\,y^{3}\cr
v_3\,y^{1}&\omega^2\, v_3\,y^{2}&\omega\,   v_3\,y^{3}
\end{pmatrix}\,.
\end{eqnarray}
\end{subequations}
%%%%%%%%%%%%%%%%%%%%%%%%%%%%%%%%%%%%%

\item[Type III. The Higgs and matter fields are $A_4$-triplets].
The mass matrix comes  from the coefficients of the product of three triplets.
For example in the Cases 3 and 5 we have
%%%%%%%%%%%%%
\begin{eqnarray}\label{eq:triplet}
      y_1^{abc}\ {\bf10_T}^a\ {\bf\B5_T}^b\ {\bf\B5_H}^c\,,
&\quad\quad&
     y_2^{abc}\ {\bf10_T}^a\ {\bf\B5_T}^b\ {\bf\B{45}_H}^c\,,
\end{eqnarray}
%
%%%%%%%%%%%%%%%%%%%%
%%%%%%%%%%%%%%%%%%%%%%%%%%%%%%%%%%%%%%%%%%%%%%%%%%%%%%%%%%%%%%%%%%%%%%%%%%
and the resulting mass matrix contributions from these operators
for the down  and charged leptons sectors are of the form:
%%%%%%%%%%%%%%%%%%%%%%%%%%%%%%%%%%%%%
\begin{eqnarray}\label{eq:tris}
\begin{pmatrix}
0              &v_3\,\gamma_2&v_2\,\gamma_1\cr
v_3\,\gamma_1  &0            &v_1\,\gamma_2\cr
v_2\,\gamma_2  &v_1\,\gamma_1&0
\end{pmatrix} ,
\end{eqnarray}
%%%%%%%%%%%%%%%%%%%%%%%%%%%%%%%%%%%%%
where $v_i$  are the vevs of the Higgs $A_4$-triplet, while the $\gamma$'s
come from the two independent $A_4$ contractions of eq. (\ref{eq:contractions}).

Similarly, in Cases 2, 3 and 5, we have contributions coming from
the operators
\begin{eqnarray}
     y_3^{abc}\ {\bf10_T}^a\ {\bf10_T}^b\ {\bf5_H}^c\,,
&\quad\quad&
     y_4^{abc}\ {\bf10_T}^a\ {\bf10_T}^b\ {\bf45_H}^c\,,
\end{eqnarray}
and the up sector mass matrix is of the form:
\begin{eqnarray}\label{eq:trisU}
\begin{pmatrix}
0            &v_3\,\gamma&\pm v_2\,\gamma\cr
\pm v_3\,\gamma  &0          &v_1\,\gamma\cr
v_2\,\gamma  &\pm v_1\,\gamma&0
\end{pmatrix},
\end{eqnarray}
where the first (second) operator gives a symmetric (an antisymmetric)
contribution.
\end{description}
%%%%%%%%%%%%%%%%%%%%%%%%%%%%%%%%%%%%%%%%%%%%%%%%%%%%%%%%%%%%%%%%%%%%%%%%%%
\subsubsection{Higgs vevs and the $A_4$ symmetry. Mixing and masses}
Let us turn now to the relation between vacuum alignment and
flavor symmetry.
The $A_4$ symmetry allows a scalar Higgs $A_4$ triplet $(H_1,H_2,H_3)$
 to get the vev in only two possible directions
(selected by the form of the Higgs potential \cite{next}).
In the so-called $S$-diagonal base, they are:
%%%%%%%%%%%%
\begin{itemize}
\item $\langle H_i\rangle=v$ for every $i$
(i.e. the direction invariant under the operator $T$);
\item $\langle H_i\rangle=v\,\delta_{i\bar k}$ for a given $\bar k$ or small
corrections around it
(i.e. the direction invariant under the operator $S$, for example
$\langle H_i\rangle=v (1,0,0)$).
\end{itemize}
%%%%%%%%%%%%
For this reason each contribution to the mass matrices
can have a defined generic form.

For Type (II) (one matter field is a $A_4$-singlet):
the mass matrices would become as:
%%%%%%%%%%%%%%%%%%%%%%%%%%%%%%%%%%
\begin{eqnarray}
M_f^{II}&\sim&
v \begin{pmatrix}
y^{1}&y^{2}&y^{3}\cr
y^{1}&y^{2}&y^{3}\cr
y^{1}&y^{2}&y^{3}
\end{pmatrix}\,
\end{eqnarray}
%%%%%%%%%%%%%
or
%%%%%%%%%%%%%
\begin{eqnarray}
M_f^{II}&\sim&
v \begin{pmatrix}
y^{1}&y^{2}&y^{3}\cr
0&0&0\cr
0&0&0
\end{pmatrix}\,
\end{eqnarray}
or similar  variants for the second form (i.e. matrices
with only one non zero row).

%%%%%%%%%%%%%%%%%%%%%%%%%%%%%%%%%%
For Type (III) (matter and Higgs fields are $A_4$-triplets):
the mass matrices would become:
%%%%%%%%%%%%%%%%
\begin{eqnarray}
M_f^{III}&\sim&
v \begin{pmatrix}
0         &\gamma_2&\gamma_1\cr
\gamma_1  &0       &\gamma_2\cr
\gamma_2  &\gamma_1&0
\end{pmatrix}
\end{eqnarray}
%%%%%%%%%%%%%
or
\begin{eqnarray}
%%%%%%%%%%%%%
M_f^{III}&\sim&
v \begin{pmatrix}
0         &0       &0\cr
0         &0       &\gamma_2\cr
0         &\gamma_1&0
\end{pmatrix},
\end{eqnarray}
%%%%%%%%%%%%%%%%
where again similar  variants for the second form are allowed (i.e. matrices
with only one non zero transposed element).

Any of the ${\bf 5_H,45_H,\B{45}_H,\overline{5}_H}$ Higgs fields appearing in the
theory which are $A_4$-triplets can have a $T$ or $S$ invariant
vev. We have a number, according to this, of the order of $\sim 2^4$ different
scenarios; we investigate the  three most promising  between them:
%%%%%%%%%%%%%%%%%%%%%%%%%%%
\begin{description}
\label{section:scenarios}
\item[Scenario {\bf (A)}:] $\langle {\B{\bf{5}}_{\bf{H}}}\rangle\propto\langle {\bf{5}_{\bf{H}}}\rangle=\{v_5,v_5,v_5\}$
 and $\langle{\B{\bf{45}}_{\bf{H}}}\rangle\propto\langle{\bf{45}_{\bf{H}}}\rangle=\{v_{45},0,0\}$,
\item[Scenario {\bf (B)}:] $\langle {\B{\bf{5}}_{\bf{H}}}\rangle\propto\langle {\bf{5}_{\bf{H}}}\rangle=\{v_5,0,0\}$
 and $\langle {\B{\bf{45}}_{\bf{H}}}\rangle\propto\langle {\bf{45}_{\bf{H}}}\rangle=\{v_{45},v_{45},v_{45}\}$,
\item[Scenario {\bf (C)}:] A mixed case with
$\langle {\B{\bf{5}}_{\bf{H}}}\rangle\propto\langle {\bf{5}_{\bf{H}}}\rangle=\{v_5,0,0\}$,
 $\langle {\B{\bf{45}}_{\bf{H}}}\rangle=\{v_{\B{45}},v_{\B{45}},v_{\B{45}}\}$, and
 $\langle {\bf{45}_{\bf{H}}}\rangle=\{v_{45},\delta v_{45},\delta v_{45}\}$.
\end{description}
%%%%%%%%%%%%%%%%%%%%%%%
As it will appear in the next sections,
scenarios {\bf(A)} and {\bf(B)}  will turn out unsatisfactory because
do not allow enough freedom to reproduce the three masses in the up sector while
Scenario {\bf(C)} will be phenomenologically interesting.
We can now write more explicit expressions for the charged fermion mass matrices for
each of these scenarios. They have the same form in all the scenarios but with
distinct parameters.

For Type (II) the mass matrices in eqs. (\ref{eq:OneIndex}) have the
form:
%%%%%%%%%%%%%%%%%%%%%%%%%%%%%%%%%%
\begin{eqnarray}\label{eq:MA4Z32}
\begin{pmatrix}
\tilde  Y^{1} &\tilde Y^{2} &\tilde  Y^{3}\cr
\tilde  y^{1} &\tilde y^{2} &\tilde  y^{3}\cr
\tilde  y^{1} &\tilde y^{2} &\tilde  y^{3}
\end{pmatrix}
\,,
\begin{pmatrix}
\tilde  Y^{1} &\tilde Y^{2} &          \tilde  Y^{3}\cr
\tilde  y^{1} &\tilde y^{2} &\omega^2\,\tilde  y^{3}\cr
\tilde  y^{1} &\tilde y^{2} &\omega\,  \tilde  y^{3}
\end{pmatrix}
\,\mbox{or}\,
\begin{pmatrix}
\tilde  Y^{1} &          \tilde Y^{2} &          \tilde  Y^{3}\cr
\tilde  y^{1} &\omega\,  \tilde y^{2} &\omega^2\,\tilde  y^{3}\cr
\tilde  y^{1} &\omega^2\,\tilde y^{2} &\omega\,  \tilde  y^{3}
\end{pmatrix}.
\end{eqnarray}

For Type (III) the mass matrix in eq. (\ref{eq:tris}) has the
form:
%%%%%%%%%%%%%%%%%%%%%%%%%%%%%%%%%%
\begin{eqnarray}%%
\label{eq:MUZ32}
M_f^{III}&\sim&
\begin{pmatrix}
\tilde  0 &\tilde \gamma_{2} &\tilde  \gamma_{1}\cr
\tilde  \gamma_{1} &0 &\tilde  \Gamma_{2}\cr
\tilde  \gamma_{2} &\tilde \Gamma_{1} &0
\end{pmatrix}.
\end{eqnarray}

The parameters are given by different expressions in each of the
vacuum alignment scenarios.
In the Scenario {\bf (A)}:
\begin{eqnarray}
&&
\begin{array}{ll}
%\tilde y^i &=& v_5\,y_{1/3}^i\,,\cr
%\tilde Y^i &=& v_5\,y_{1/3}^i + c\,v_{45}\,y_{2/4}^i\,,
\tilde y^i = \tilde y^i(v_{\B5}),\quad & \tilde Y^i = \tilde Y^i(v_{\B5},\,v_{\B{45}});\\
\tilde \gamma_i = \tilde \gamma_i(v_{\B5}),\quad &  \tilde \Gamma_i
=\tilde \Gamma_i(v_{\B5},\,v_{\B{45}});
\end{array}
\end{eqnarray}
in Scenarios {\bf (B)}, {\bf(C)}:
\begin{eqnarray}
&&
\begin{array}{ll}
%\tilde y^i &=& v_{45}\,y_{2/4}^i\,,\cr
%\tilde Y^i &=& v_5\,y_{1/3}^i + c\,v_{45}\,y_{2/4}^i\,,
\tilde      y^i = \tilde y^i(v_{\B{45}}),\quad & \tilde      Y^i = \tilde Y^i(v_{\B5},\,v_{\B{45}});\\
\tilde \gamma_i = \tilde \gamma_i(v_{\B{45}}),\quad & \tilde \Gamma_i = \tilde \Gamma_i(v_{\B5},\,v_{\B{45}})\,.
\end{array}
\end{eqnarray}
%
%%%%%%%%%%%%%%%%%%%%%%%%%%%%%%%%%%%
The first mass matrix in eq. (\ref{eq:MA4Z32}) have a zero eigenvalue
and cannot be compatible with the experimental data.
Only the second and the third mass matrix in eq. (\ref{eq:MA4Z32}) have three non
zero eigenvalues.
We conclude  that in our models there is at least a viable $A_4$ flavor solution when
the field
$\bf{\B5_T}$ composes of three singlets and they do not transform equivalently.

We will focus now in the
 mass matrix in eq. (\ref{eq:MUZ32}).
This matrix has the attractive feature
of having three independent eigenvalues. We will study
 it in two opposite limits.
This matrix is diagonalized in general by left and right matrices
as $M_f^{III}=V M_f^{diag} W^\dagger$.
First, in the limit $\tilde \Gamma_i \rightarrow \tilde\gamma_i$,
the right and left mixing matrices become the same:
\begin{eqnarray}\label{eq:MixingA4Z32}
V\sim W&=&
\begin{pmatrix}
1 &1 &1\cr
1 &\omega  &\omega^2\cr
1 &\omega^2&\omega
\end{pmatrix}
\,,
\end{eqnarray}
(for the charged lepton the mixing matrices are the Hermitian conjugate of the quark ones)
and the masses are given by (in the same limit):
\begin{equation}\label{eq:gamma}
M_f^{diag}=diag
(
\tilde\gamma_1+\tilde\gamma_2\,,
\tilde\gamma_2+\omega\,\tilde\gamma_1\,,
\tilde\gamma_2+\omega^2\,\tilde\gamma_1 )\,.
\end{equation}
On the other side, in the limit $\tilde\Gamma_i\gg \tilde\gamma_i$,
one can use on the right side the mixing matrix
\begin{eqnarray}\label{eq:MixingZ2}
W&=&\begin{pmatrix}
1 & 0 &0\cr
0 & 0 &1\cr
0 & -1 &0\cr
\end{pmatrix},
\end{eqnarray}
and the three eigenvalues, or masses, are
$(0,|\tilde\Gamma_1|,|\tilde\Gamma_2|)$.
\section{
Flavor predictions in the quark mixing and
Fits of the charged fermion masses}

\subsection{Considerations for cases 2 and 4}
{\em Cabibbo angle eliminates Case 4:}
First of all we observe that in Case 4 all the charged fermion masses
are exactly diagonal, so they cannot generate the CKM matrix at tree level. This is
an enough reason for considering this
case  of no phenomenological interest and it  will not be discussed any further.

{\em In Case 2 charged fermion masses require that the $\bf{\B5_T}$ transforms
differently each other:} the mass matrices $M_D$ and $M_E$
are of the Type (III) discussed before, they  originate from
coupling constants associated to operators where
{\em the Higgs is a triplet and one kind of fermion fields contains three $A_4$-singlets},
i.e. they are of the form in eq. (\ref{eq:MA4Z32}). The mass matrix $M_U$ is
originated from a {\em $A_4$-triplet Higgs}, i.e. it
is of the form in eq. (\ref{eq:MUZ32}).
As discussed in the sec. {\bf\ref{sec:Pred}}, the mass matrices $M_D$ and $M_E$
have three non zero eigenvalues only if the three $\bf{\B5_T}$ transforms as
1, 1', and 1''.

As already mentioned, the mass and quark mixing matrices in this sub-case has
been well studied in literature
\cite{Mitra:2008bn,Luhn:2008sa,Albright:2008qb,Parida:2008pu,Lin:2008aj,Adhikary:2008au,Plentinger:2008nv,Ding:2008rj,Plentinger:2008up,Bazzocchi:2008rz}
in other phenomenological scenarios and their detailed study in our concrete scenario
will appear elsewhere \cite{next}.

\subsection{Case 5: realistic charged fermion masses and quark mixing}\label{sec:cases3}
Let us focus now in the most phenomenologically attractive case.
The  Case 5 includes the following  explicit terms:
\begin{subequations}\label{eq:lagr5}
\begin{eqnarray}
W_0&=&
      Y_1^{abc}\ {\bf10_T}^a\ {\bf\B5_T}^b\ {\bf\B5_H}^c
    + Y_2^{abc}\ {\bf10_T}^a\ {\bf\B5_T}^b\ {\bf\B{45}_H}^c
\nonumber\\&&\quad\quad
    + Y_3^{abc}\ {\bf10_T}^a\ {\bf10_T}^b\ {\bf5_H}^c
    + Y_4^{abc}\ {\bf10_T}^a\ {\bf10_T}^b\ {\bf45_H}^c,\\
W_1&=&
      \gamma^{abc}\ {\bf\B5_T}^a\ {\bf24_T}^b\ {\bf45_H}^c
    + \beta^{abc}\ {\bf\B5_T}^a\ {\bf24_T}^b\ {\bf5_H}^c,\\
W_2&=&
      m_\Sigma\ Tr ({\bf24_H\ 24_H})
    + \lambda_\Sigma\ Tr({\bf24_H\ 24_H\ 24_H}) +
\nonumber\\&&\quad\quad
       m\ {\bf24_T}^a\ {\bf24_T}^a
     + \lambda\ {\bf24_T}^a\ {\bf24_T}^a\ \bf{24_H},\\
W_3&=&
  \lambda_H\ {\bf\B5_H}^a\ {\bf24_H}\ {\bf5_H}^a
      + c_H\ {\bf\B5_H}^a\ {\bf24_H}\ {\bf45_H}^a
      + b_H\ {\bf\B{45}_H}^a\ {\bf24_H}\ {\bf5_H}^a
\nonumber\\&&\quad\quad
    + m_{5}\ {\bf\B5_H}^a\ {\bf5_H}^a
   + m_{45}\ {\bf\B{45}_H}^a\ {\bf45_H}^a
      + a_H\ {\bf\B{45}_H}^a\ {\bf45_H}^a\ {\bf24_H}.
\end{eqnarray}
\end{subequations}
The mass matrices
for $M_D$, $M_E$, and $M_U$ are given here by coupling constants associated to
{\em $A_4$-triplet Higgs } operators and the charged fermion masses
are of the form as in eq. (\ref{eq:MUZ32}). As we will show below,
there is room for three independent
eigenvalues and the phenomenological observed charged fermion hierarchy
can be easily reproduced.

Let us write in detail the operators
$y_1\, {\bf10_T}\,{\bf\B5_T}\,{\bf\B5_H}$ and $y_2\, {\bf10_T}\,{\bf\B5_T}\,{\bf\B{45}_H}$
which generate both the down and charged lepton mass matrices.
Under SM gauge group invariance, with the field notation given
in \cite{Slansky:1981yr}, the operators read as:
\begin{subequations}
\begin{eqnarray}
y_1\, {\bf10_T}\,{\bf\B5_T}\,{\bf\B5_H} &\rightarrow&
{\bf Q_i^{\alpha}} M_{ij} {\bf d_j} {\bf {\B H}^{\alpha}} +{\bf e_i} M_{ij} {\bf L_j^\alpha} {\bf \B H^{\alpha}};
\\
y_2\, \bf{10_T}\,\bf{\B5_T}\,\bf{\B{45}_H} &\rightarrow&
{\bf Q}_i^\alpha (2\,M_{ij}) {\bf d}_j {\bf\B H}_{(1,2)}^\alpha +
{\bf e}_i (-6\,M_{ij}) {\bf L}_j^\alpha {\bf\B H}_{(1,2)}^\alpha\,,
\end{eqnarray}
\end{subequations}
where  $\alpha$ is the $SU(2)$ index.
The two other operators
$
    y_3\, {\bf10_T}\,{\bf10_T}\,{\bf5_H}$ and
    $y_4\, {\bf10_T}\,{\bf10_T}\,{\bf45_H}
$
generate the up mass matrix in a similar way; in particular the
operator proportional to ${\bf5_H}$ generate a symmetric
term, while the operator proportional to
${\bf45_H}$ generate an antisymmetric term:
\begin{subequations}
\begin{eqnarray}
y_3\, {\bf10_T}\,{\bf 10_T}\,{\bf 5_H} &\rightarrow&
{\bf Q}_i^\alpha M_{ij} {\bf u}_j {\bf H}^\alpha +
{\bf u}_i M_{ij} {\bf Q}_j^\alpha {\bf H}^\alpha\,;
\\
y_4\, {\bf10_T}\,{\bf 10_T}\,{\bf45_H} &\rightarrow&
{\bf Q}_i^\alpha M_{ij} {\bf u}_j {\bf H}_{(1,2)}^\alpha
 -{\bf u}_i \,M_{ij} {\bf Q}_j^\alpha {\bf H}_{(1,2)}^\alpha\,.
\end{eqnarray}
\end{subequations}
We consider next  the three possible scenarios with respect to
the Higgs vacuum alignments in the $A_4$ structure we mentioned previously and
write in detail the mass matrices in each of them.

\subsubsection{Matrices for Scenarios {\bf (A)} and {\bf (B)}}

In Scenario {\bf (A)} we impose
 $\langle {\B{\bf{5}}_{\bf{H}}}\rangle\propto\langle {\bf{5}_{\bf{H}}}\rangle=\{v_5,v_5,v_5\}$ and $\langle{\B{\bf{45}}_{\bf{H}}}\rangle\propto\langle{\bf{45}_{\bf{H}}}\rangle=\{v_{45},0,0\}$. Under these assumptions
the mass matrices in
%%%%%%%%%%%%%%%%%%%%%%%%%%%%%%%%%%%%%%%%%%%%%%%%%%%%%%%%%%%%
the up sector are too constrained. In particular there are only two free
parameters %(remember that $\tilde\Gamma_u^2=2\tilde\gamma_u-\tilde\Gamma_u^1$)
and it is not possible to fit the huge hierarchy among the masses.

In Scenario {\bf (B)}, the vacuum expectation values are as
 $\langle {\B{\bf{5}}_{\bf{H}}}\rangle\propto\langle {\bf{5}_{\bf{H}}}\rangle=\{v_5,0,0\}$ and
$\langle{\B{\bf{45}}_{\bf{H}}}\rangle\propto\langle{\bf{45}_{\bf{H}}}\rangle=\{v_{45},v_{45},v_{45}\}$.
Under the assumptions
$v_{\B5},v_{\B45}\gg v_{45}$ and $\gamma^1_1\gg\gamma^2_1$
we obtain that the 3rd generation is much more heavy than the other two, 
the relation $\tilde\Gamma_e^1\simeq \tilde\Gamma_d^1$
holds and we get the bottom-tau unification.
However, also in this scenario the up sector mass
matrix is too constrained to fit the quark masses in detail.
In particular, as in scenario {\bf (A)}, there are only two free
parameters %(remember that $\tilde\Gamma_u^2=2\tilde\gamma_u-\tilde\Gamma_u^1$)
and it is not possible at all to fit the three quark masses.

\subsubsection{Scenario {\bf (C)}: $\langle {\B{\bf{5}}_{\bf{H}}}\rangle\propto\langle {\bf{5}_{\bf{H}}}\rangle=\{v_5,0,0\}$,
$\langle{\B{\bf{45}}_{\bf{H}}}\rangle=\{v_{\B{45}},v_{\B{45}},v_{\B{45}}\}$,
and $\langle{\bf{45}_{\bf{H}}}\rangle=\{v_{45},\delta v_{45},\delta v_{45}\}$}\label{sec:scenarioC}
In this scenario the down and charged lepton mass matrices are given by:
\begin{subequations}\label{eq:matrix2}
\begin{eqnarray}
\label{eq:Dmatrix2}
M_{D}&=&
\begin{pmatrix}
         0 & 2\gamma_{2}^{1}v_{\B{45}} & 2\gamma_{2}^{2}v_{\B{45}} \\
         2\gamma_{2}^{2}v_{\B{45}} & 0 &
2\gamma_{2}^{1}v_{\B{45}}+\gamma_{1}^{1}v_{\B{5}} \\
         2\gamma_{2}^{1}v_{\B{45}} &
2\gamma_{2}^{2}v_{\B{45}}+\gamma_{1}^{2}v_{\B{5}} & 0
       \end{pmatrix}
\quad\equiv\quad\begin{pmatrix}
         0 & \tilde\gamma_{d}^{1} & \tilde\gamma_{d}^{2} \\
         \tilde\gamma_{d}^{2}& 0 & \tilde\Gamma_{d}^{1} \\
         \tilde\gamma_{d}^{1}&\tilde\Gamma_{d}^{2}&0
\end{pmatrix},
\\
\label{eq:Ematrix2}
M_{E}&=&
       \begin{pmatrix}
         0 & -6\gamma_{2}^{2}v_{\B{45}} & -6\gamma_{2}^{1}v_{\B{45}} \\
         -6\gamma_{2}^{1}v_{\B{45}} & 0 &
-6\gamma_{2}^{2}v_{\B{45}}+\gamma_{1}^{2}v_{\B{5}} \\
         -6\gamma_{2}^{2}v_{\B{45}} &
-6\gamma_{2}^{1}v_{\B{45}}+\gamma_{1}^{1}v_{\B{5}} & 0
       \end{pmatrix}
\quad\equiv\quad\begin{pmatrix}
         0 & -3\tilde\gamma_{d}^{2} & -3\tilde\gamma_{d}^{1} \\
         -3\tilde\gamma_{d}^{1}& 0 & \tilde\Gamma_{e}^{2} \\
         -3\tilde\gamma_{d}^{2}&\tilde\Gamma_{e}^{1}&0
        \end{pmatrix},
\end{eqnarray}
\end{subequations}
where we introduced the short hand notations ($i=1,2$):
\begin{subequations}
\begin{eqnarray}\label{eq:DCoeff2}
\tilde\gamma_d^i&=&2\gamma_2^i\,v_{\B{45}},\\
\tilde\Gamma_d^i&=&\gamma_{1}^{i}\,v_{\B{5}}+2\gamma_{2}^{i}\,v_{\B{45}},\\
\tilde\Gamma_e^i&=&\gamma_{1}^{i}\,v_{\B{5}}-6\gamma_{2}^{i}\,v_{\B{45}}\label{eq:ECoeff2}.
\end{eqnarray}
\end{subequations}
%%%%%%%%%%%%%%%%%%%%%%%%%%%%%%%%%%%%%%%%%%%%%%%%%%%%%%%%%
A relation between the masses for charged leptons and down quarks
is given by:
\begin{eqnarray}
\label{eq:mdme}
M_D-M_E^t&=& v_{\B{45}} Y_2,
\end{eqnarray}
where the ``effective'' Yukawa matrix $Y_2$ is given by:
\begin{eqnarray}
Y_2&=& 8
\begin{pmatrix}
         0 & \gamma^1_2 & \gamma^2_2 \\
         \gamma^1_2 &0& \gamma^1_2 \\
         \gamma^2_2 & \gamma^2_2 & 0 \\
\end{pmatrix}.
\end{eqnarray}
The relation given by eq.(\ref{eq:mdme}) is a particular case of the D-E relation in
SUSY $SU(5)$ theories. In our case the texture of the ``Yukawa'' matrix $Y_2$ is
predicted from the $A_4$ symmetry and its coefficients are given by the
$A_4$ coupling constants.
If we want to keep the bottom-tau unification at GUT scale, the matrix
$Y_2$ must only modify the relation between first and second quark and lepton masses.

Other matrix relations can be easily obtained.  One gets directly
 from the expressions of matrices $M_D,M_E$:
\begin{eqnarray}
\label{eq:mdme2} 3 M_D+M_E^t&=& v_{\B{5}} Y_5,
\end{eqnarray}
where the  matrix $Y_5$ is given by:
\begin{eqnarray}
Y_5&=& 4
\begin{pmatrix}
         0 & 0 & 0\\
         0&0& \gamma^1_1 \\
         0 & \gamma^2_1 & 0 \\
\end{pmatrix}.
\end{eqnarray}
In the limit $v_5\sim 0$ we obtain, instead of
$M_D\sim M_E^t$ in minimal $SU(5)$, the relation:
$$ M_D\sim -M_E^t/3\,.$$.
%%%%
It is possible to obtain also some  useful ``sum rules''  for the
squared masses of quarks and charged leptons by taking the
trace of the matrices $9 M_D M_D^{\dagger}\pm M_E M_E^{\dagger} $:
\begin{subequations}\label{eq:mdmeX}
\begin{eqnarray}
\label{eq:mdme3} 9 \sum_{down} m_q^2-\sum_{l} m_l^2&=& 48 Re (\gamma^1_1 \gamma^{1\star}_2+
\gamma^2_1\gamma^{2\star}_2 ) v_{\B{45}}v_{\B{5}}+8 (\mid \gamma^1_1\mid^2+\mid \gamma^2_1\mid^2) v_{\B{5}}^2 ,\\
\label{eq:mdme4} 9 \sum_{down} m_q^2+\sum_{l} m_l^2&=&
216 (\mid \gamma^1_2\mid^2+\mid \gamma^2_2\mid^2) v_{\B{45}}^2 +B v_{\B{45}}v_{\B{5}}+C v_{\B{5}}^2,
\end{eqnarray}
\end{subequations}
where $A,B$ are simple expressions depending on the $\gamma^i_j$ constants. Note
that the first sum rule, eq.(\ref{eq:mdme3}) would become specially simple if the
the $2\times 2$ matrix of constants $(\gamma^{i}_j)$ would be unitary. In such a case the term
proportional to $v_{45}$ would vanish. Expressions in eqs. (\ref{eq:mdmeX})
will prove themselves to be useful in the numerical fits of the next section.
Let us proceed now to the up quark mass matrix. In this scenario
we obtain  the following form:
\begin{eqnarray}
\label{eq:Umatrix3}
M_{U}&=&
\begin{pmatrix}
         0 & -8\left(\gamma_{4}^{1}-\gamma_{4}^{2}\right)\delta v_{45} &
 8\left(\gamma_{4}^{1}-\gamma_{4}^{2}\right)\delta v_{45} \\
        8\left(\gamma_{4}^{1}-\gamma_{4}^{2}\right)\delta v_{45} & 0 &
 -8\left(\gamma_{4}^{1}-\gamma_{4}^{2}\right)v_{45}
 +4\left(\gamma_{3}^{1}+\gamma_{3}^{2}\right)v_{5}
\\
         -8\left(\gamma_{4}^{1}-\gamma_{4}^{2}\right)\delta v_{45} &
  8\left(\gamma_{4}^{1}-\gamma_{4}^{2}\right)v_{45}
 +4\left(\gamma_{3}^{1}+\gamma_{3}^{2}\right)v_{5}
& 0 \\
       \end{pmatrix}
\nonumber\\
&&\quad\equiv\quad\begin{pmatrix}
         0 & -\tilde\gamma_{u} & \tilde\gamma_{u} \\
         \tilde\gamma_{u}& 0 & \tilde\Gamma_{u}^{1} \\
         -\tilde\gamma_{u}&\tilde\Gamma_{u}^{2}&0
\end{pmatrix},
\end{eqnarray}
where we have defined the following coefficients:
\begin{subequations}
\begin{eqnarray}
\tilde\gamma_u&=&8\left(\gamma_{4}^{1}-\gamma_{4}^{2}\right)\delta v_{45},\\
\tilde\Gamma_u^1&=&4\left(\gamma_{3}^{1}+\gamma_{3}^{2}\right)v_{5}-8\left(\gamma_{4}^{1}-\gamma_{4}^{2}\right)v_{45},\\
\tilde\Gamma_u^2&=&4\left(\gamma_{3}^{1}+\gamma_{3}^{2}\right)v_{5}+8\left(\gamma_{4}^{1}-\gamma_{4}^{2}\right)v_{45}.
\end{eqnarray}
\end{subequations}

It is straightforward to derive a sum rule for the up squared masses from the
trace of the matrix $M_U M_U^{\dagger}$; we obtain:
\begin{eqnarray}
\label{eq:mu}
\sum_{up} m_q^2& =& \mid \tilde\gamma_u \mid^2+\mid \tilde\Gamma_u^1 \mid^2+\mid \tilde\Gamma_u^2 \mid^2,
\end{eqnarray}
which reduces quickly to the expected sum of the squares of the three vevs.
The matrix $M_U$  turns out to be phenomenologically viable in this case.
We can reproduce the three up masses
by imposing the hierarchy
$\tilde\Gamma^{1}\gg\tilde\Gamma^{2}\gg\tilde\gamma$, i.e. with a fine tuning
in $\tilde\Gamma^2$, and the eigenvalues of the matrix are then approximately:
\begin{eqnarray}
\{m_u,m_c,m_t\}&\simeq&\left\{\left|\frac{(\tilde\gamma_u)^2}{\tilde\Gamma_u^2}\right|,
\left|\tilde\Gamma_u^2\right|,
\left|\tilde\Gamma_u^1\right|
\right\}
\end{eqnarray}
and the experimental values can be easily accommodated.

\subsubsection{Numerical Masses and CKM  in Case 5, scenario {\bf (C)}. }
Information about the free parameters of the model can be obtained by performing
a fit to the experimental values of the masses of up and down quarks and charged
leptons at the GUT scale.
For our purpose it will be enough to consider
the run quark and lepton masses of \cite{Das:2000uk}.
The run masses, for example
by assuming an unification scale $\mu=2\times 10^{16}$ GeV,
a SUSY scale $M_S=1\ TeV$ and $\tan\beta(M_S)=10$,
are given in table (\ref{tab:masses}) (from \cite{Das:2000uk}).
\begin{table}[hbt]
\begin{tabular}{|l|l|l|}\hline
$m_u$=  $0.72^{+0.14}_{-0.15}$& $m_c$=  $210^{+19}_{-21}$& $m_t$= $82^{+30}_{-15}\times 10^3$\\ \hline
$m_d$=$1.5^{+0.4}_{-0.2}$& $m_s$=$29\pm4$&   $m_b$= $1060^{+140}_{-90}$\\ \hline
$m_e$= $0.3585\pm 0.0003$&$m_\mu$=  $75.67\pm 0.05$&$m_\tau$=$1292.2^{+1.3}_{-1.2}$\\ \hline
\end{tabular}
\caption{Running masses (MeV) for unification scale $\mu=2\times 10^{16}$ GeV,
 SUSY scale $M_{S}=1\ TeV$ and $\tan\beta(M_S)=10$.}
\label{tab:masses}
\end{table}
We have computed the exact eigenvalues of our mass matrices,
eqs. (\ref{eq:Dmatrix2}),(\ref{eq:Ematrix2})
and (\ref{eq:Umatrix3}). The best fit four parameters
$\tilde\gamma_d^i$, $\tilde\Gamma_d^i$ have been obtained
 by  a standard $\chi^2$ minimization of  the error-weighted  distance
from the theoretical eigenvalues to running masses of table (\ref{tab:masses})
computed from low scale experimental values.

The resulting parameters
in the down and charged lepton sectors,
are:
%%%%%%%%%%%%%%%%%%%%%%%%%%%%%%%%%%%%%%%%%%%%%%%
\begin{eqnarray}
\tilde\Gamma_d^2 = (1277.9 - 0.0145\ i)\ MeV
&,\quad&
\tilde\Gamma_d^1 = (-26.83 - 0.494\ i)\ MeV,\\
\tilde\gamma_d^1 = (11.849 - 1.069\ i)\ MeV
&,\quad&
\tilde\gamma_d^2 = (-3.32 + 0.37\ i)\ MeV.
\end{eqnarray}
The fit is acceptable,  $\chi^2/ndf\sim 1.5$ and the pull out of the best fit
masses obtained in return is small:
\begin{eqnarray}
\{m_d,m_s,m_b\}_{best\ fit}&=&\{0.3, 27, 1280\}\ MeV,\\
\{m_e,m_\mu,m_\tau\}_{best\ fit}&=&\{0.359, 75.67, 1292.2\}\ MeV\,.
\end{eqnarray}
We notice that there is a preference for the mass $m_d$ to be lower
and  $m_b$  to be higher than expected.
%%%%%%
Additional information about the size of the free parameters and vevs of the model
can be obtained by using the sum rules in eqs. (\ref{eq:mdmeX}).
The experimental value for the quantities:
\begin{subequations}
\begin{eqnarray}
 9 \sum_{down} m_q^2-\sum_{l} m_l^2&=& (2.9\pm 0.1\ GeV)^2 \label{eq:sum1},\\
 9 \sum_{down} m_q^2+\sum_{l} m_l^2 &=& (3.4\pm 0.1\ GeV)^2\label{eq:sum2},
\end{eqnarray}
and the quotient:
\begin{eqnarray}
\left( 9 \sum_{down} m_q^2-\sum_{l} m_l^2\right)/
\left( 9 \sum_{down} m_q^2+\sum_{l} m_l^2\right)&\simeq& 0.7\ (exp.),
\end{eqnarray}
\end{subequations}
 can be used to extract information about the $v_5,v_{45}$ vevs. %using
%%%%%%%%%
The three parameters of the  up sector $\tilde\gamma_u$ and $\tilde\Gamma_u^{i=1,2}$ are obtained numerically directly from
the three run masses and the result is:
\begin{eqnarray}
\tilde\Gamma_u^1 &=& 82.0\, MeV, \quad \tilde\Gamma_u^2 = 0.20\, MeV,
\quad\tilde\gamma_u = 0.02\, MeV.
\end{eqnarray}
These numerical values are in agreement with the
expectations,  $\tilde\gamma_u \propto \delta v_{45}$ and it is much smaller.
Moreover we can write, directly from the definitions:
\begin{eqnarray}
\left|
\frac{\tilde\gamma_u}
{\tilde\Gamma_u^2-\tilde\Gamma_u^1}\right|&=&\frac{1}{2}\frac{\delta v_{45}}{v_{45}};
\end{eqnarray}
this means, using the fitted values above:
\begin{eqnarray}
\frac{\delta v_{45}}{v_{45}} =5\times 10^{-4}.
\end{eqnarray}
This estimation can be considered a direct fit to the experimental masses.

We can estimate the CKM at this stage.  Any  mass matrix can be diagonalized as:
\begin{eqnarray}\label{eq:diag}\label{eq:mixdef}
M_U=U_U\, M_U^{diag}\, V_U^\dagger\,,\quad&
M_D=U_D\, M_D^{diag}\, V_D^\dagger\,,&\quad
M_E=U_E\, M_E^{diag}\, V_E^\dagger,
\end{eqnarray}
from them we can obtain the usual CKM matrix as
\begin{eqnarray}\label{eq:CKM}
U_{CKM}= U_U^\dagger\, U_D\,.
\end{eqnarray}
We first  build the matrices $M_U,M_E,M_D$ from the best fit of the
parameters obtained above and
then numerically find the diagonalizing matrices.
The resulting CKM mixing matrix is obtained from eq.(\ref{eq:CKM}):
\begin{eqnarray}
|U_{CKM}|&=&\begin{pmatrix}
0.984& 0.180&0.0091\\
0.180&0.984&0.0006\\
0.009&0.001&0.9999
\end{pmatrix}.
\end{eqnarray}
We notice that our model reproduce the main features of the CKM structure,
i.e. a relatively large Cabibbo angle and very small
$(1,3)$ and $(3,1)$ entries.
Other characteristics as the size of the entries $(2,3)$ and $(3,2)$ might be
improved by performing a more refined global fit
of masses and CKM mixing matrix together which is beyond the scope of this work.

\section{Neutrino mass and lepton mixing matrices}
We investigate now the neutrino mass matrices, their masses
 and the resulting PMNS mixing matrix. We know,
as conclusion of the study of the charged fermion masses,  that from
the five initial Cases only two, the 3 and 5, remain phenomenologically
interesting. The vacuum alignment scenario {\bf C} is needed in both cases.

Let us present first  a summary of  the results of the Case 3.
Here the adjoint matter field $\bf24_T$ is singlet under the $A_4$
symmetry while all the  Higgs fields are triplets, except ${\bf 24_H}$
according to table (\ref{tab:A4all}).
The neutrino masses corresponding to  the scenarios  {\bf (A)} and {\bf (B)}
(see sec. \ref{section:scenarios})
are quite general, only one of them is zero.
As a general feature, the neutrino masses can be accommodated well in
both scenarios.
Moreover since $U_E$ is close to the identity, it turns out that
the observed tribimaximal
mixing structure of PMNS lepton matrix can be obtained ligated to the existence
of an inverted neutrino hierarchy.
%We report the full computation in Appendix \ref{app:numasscase3} %% version 1.
The situation is not the same in scenario {\bf (C)}.
In this scenario the neutrino hierarchy results to be the normal one and
as a consequence the  desired lepton mixing matrix cannot be obtained.
Scenario {\bf (C)}, as explained in the previous section, is the only one which
 reproduce the up sector masses.
We arrive to the conclusion that, for a $A_4$-singlet $\bf24_T$,
  is impossible at the same time to reproduce charged fermion and neutrino masses.

We will study next, now in detail, the much more interesting Case 5.

\subsection{Neutrino masses in Case 5: $\bf24_T$  is a  $A_4$-triplet.}
Here all the the Higgs fields
$\bf\B5_H$, $\bf\B{45}_H$, $\bf5_H$, $\bf45_H$ are
flavor triplets (table \ref{tab:A4all}).
Once the Higgs multiplets acquire a vev, the quadratic
mass term between the $\bf\B5_T$ and the $\bf24_T$ fields,
i.e. those originating from
$W_1=\gamma\, {\bf\B5_T}\,{\bf24_T}\,{\bf45_H}+ \beta\, {\bf\B5_T}\,{\bf24_T}\,{\bf5_H},$
becomes of the form:
\begin{subequations}
\begin{eqnarray}
{\bf\B5}^a M_{ab} {\bf24}^b,
\end{eqnarray}
where the mass matrix is given by:
\begin{eqnarray}
M=(M_{ab}) &=&\begin{pmatrix}
0&\gamma_1 v_{45}^3 + \beta_1 v_5^3&\gamma_2 v_{45}^2 + \beta_2 v_5^2\\
\gamma_2 v_{45}^3 + \beta_2 v_5^3&0&\gamma_1 v_{45}^1 + \beta_1 v_5^1\\
\gamma_1 v_{45}^2 + \beta_1 v_5^2&\gamma_2 v_{45}^1 + \beta_2 v_5^1&0\\
\end{pmatrix},
\end{eqnarray}
\end{subequations}
where the $A_4$ coupling constants $\gamma_i,\beta_i$ are defined as before.

The SM neutrino masses are generated through  a type I+III SeeSaw mechanism
mediated by the $SU(2)_{weak}$  singlet and the neutral component of
$SU(2)_{weak}$ triplet   in the $\bf24_T$ matter field, respectively
$\rho_0,\rho_3$ with masses
$M_{\rho_0},M_{\rho_3}$ (see \cite{Slansky:1981yr} for the SM decomposition of the fields in our model).
The neutrino mass matrix is of the form:
\begin{subequations}\label{eq:nu}
\begin{eqnarray}\label{eq:nu5}
M_\nu&=&
\frac{1}{M_{\rho_3}}\begin{pmatrix}
a_{13}^2 +a_{22}^2&a_{22} a_{11}&a_{13} a_{21}\\
a_{22} a_{11}&a_{23}^2 +a_{11}^2&a_{23} a_{12}\\
a_{21} a_{13}&a_{12} a_{23}&a_{12}^2 +a_{21}^2
\end{pmatrix}
+
\frac{1}{M_{\rho_0}}\begin{pmatrix}
b_{13}^2+b_{22}^2&b_{22} b_{11}&b_{13} b_{21}\\
b_{22} b_{11}&b_{23}^2+b_{11}^2&b_{23} b_{12}\\
b_{21} b_{13}&b_{12} b_{23}&b_{12}^2+b_{21}^2
\end{pmatrix},
\end{eqnarray}
where ($p=1,2,j=1,2,3$):
\begin{eqnarray}
a_{pj}&=&-3\gamma_p v_{45}^j + \beta_pv_5^j, \\
b_{pj}&=&\frac{\sqrt{15}}{2}\left(\gamma_p v_{45}^j + \frac{1}{5}\beta_pv_5^j\right).
\end{eqnarray}
\end{subequations}
%%%%%%%%%%%%%%%%%%%%%%%%%%%%%%%%%%%%%%%%%%%%%%%%%%%%%%%%%%%%%%%%%%%%%%%%%%%%
We assume that the vevs are in the following directions (Scenario {\bf (C)}):
$v_5^j=\delta_{j1} v_5=(1,0,0) v_5$
and $v_{45}^j=(1,\epsilon,\epsilon )\, v_{45}$, we get:
\begin{subequations}
\begin{eqnarray}
a_{p1}&=&-3\gamma_p v_{45} + \beta_pv_5,\\
a_{pj}&=&-3\,\epsilon\,\gamma_p v_{45}\quad (j=2,3),\\
b_{p1}&=&\frac{\sqrt{15}}{2}\left(\gamma_p v_{45}+ \frac{1}{5}\beta_pv_5\right),\\
b_{pj}&=&\frac{\sqrt{15}}{2}\,\epsilon\,\gamma_p v_{45}\quad (j=2,3).
\end{eqnarray}
\end{subequations}
With these parameters
the low energy neutrino mass matrix is given by:
\begin{eqnarray}
M_\nu&=&
\begin{pmatrix}
\epsilon^2 (A_{11}+A_{22})
&\epsilon\,(B_{21}+A_{21})&\epsilon\,(B_{12}+A_{12})\\
\epsilon\,(B_{21}+A_{21})&\epsilon^2\,A_{22}+A_{11}+2B_{11}+C_{11}
  &\epsilon^2\,A_{12}\\
\epsilon\,(B_{12}+A_{12})&\epsilon^2\,A_{12}
  &\epsilon^2\,A_{11}+A_{22}+2B_{22}+C_{22}
\end{pmatrix},
\label{eq:mnu11}
\end{eqnarray}
where we have considered explicitly the dependence on  $\epsilon$ of each entry,
 the $A$'s and $B$'s are simple combinations of the $a_{pj},b_{pj}$ parameters.

Our objective is to study under which conditions the matrix (\ref{eq:mnu11})
is diagonalized by a tribimaximal unitary matrix. Let us note that for
a generic matrix,  with arbitrary $a,b,c$ coefficients:
\begin{eqnarray}\label{eq:muA4}
M^{TBM}=\begin{pmatrix}
c&b&b\\
b&c+a&b-a\\
b&b-a&c+a
\end{pmatrix},
\end{eqnarray}
the diagonalizing matrix
($M^{TBM}=U_\nu M^d U_\nu^\dagger$), as a matter of fact, is the tribimaximal matrix
\cite{Harrison:2002er}:
\begin{eqnarray}\label{eq:TBM}
U_\nu^{TBM}&=&
\begin{pmatrix}
 \sqrt{\frac{2}{3}}  &\frac{1}{\sqrt{3}}  & 0 \cr
 -\frac{1}{\sqrt{6}} &\frac{1}{\sqrt{3}}  &\frac{1}{\sqrt{2}}\cr
 -\frac{1}{\sqrt{6}} &\frac{1}{\sqrt{3}}  &-\frac{1}{\sqrt{2}}
\end{pmatrix}.
\end{eqnarray}
%%%%%%%%%%%%%%
The PMNS lepton mixing matrix is given by:
\begin{eqnarray}
U_{PMNS}&=&U_E^\dagger U_\nu
\end{eqnarray}
where $U_E$ is defined in eq. (\ref{eq:mixdef}) and $U_\nu$ is defined
by:
\begin{eqnarray}
	M_\nu&=&U_\nu M_\nu^{diag} U_\nu^t\,.
\end{eqnarray}
We conclude from here that if we impose the condition that the matrix
(\ref{eq:mnu11})
is of the form of the matrix
(\ref{eq:muA4})
the resulting lepton mixing %($U_{PMNS}=U_\nu^\dagger U_e$ )
will be
compatible with the experimental evidence \cite{Chauhan:2006im},
since we know from the previous section that,
 in this case and with this vev scenario, the charged lepton matrix $U_E$  is
a small perturbation of the same order of CKM mixing matrix.
We thus arrive to a ``complementarity'' relation $U_{PMNS}\approx U_{\nu}^{TBM}\times U_{CKM}$
\cite{Chauhan:2006im}.
%%%%%%%%%%%%%%%%%%%%%%%%%%%%%%%%%%%%%%
The condition $M_\nu\sim M^{TBM}$ implies the following equalities for the
coefficients:
\begin{eqnarray}
M_{\nu,12}&=& M_{\nu,13},\quad
M_{\nu,22}=M_{\nu,33},\\
M_{\nu,22}+M_{\nu,23}&=&M_{\nu,11}+M_{\nu,12}.
\end{eqnarray}
A solution for these equations is given,
at leading order in $\epsilon$, by:
\begin{subequations}
\begin{eqnarray}
\beta_2&=&\beta_1 \epsilon f_1(\alpha), \\
 \gamma_2&=&\frac{v_5}{v_{45}} \beta_1 \epsilon f_2(\alpha),\\
 \gamma_1&=&\frac{v_5}{v_{45}}
 \beta_1 f_3(\alpha).
\end{eqnarray}
\end{subequations}
where the parameter $\beta_1$ is left free,
$\alpha=M_{\rho_3}/M_{\rho_0}$ and
$f_1=
\frac{16\sqrt{\alpha}(4 i \sqrt{15} + 16\sqrt{\alpha} - i \sqrt{15}\alpha)}
{(20 + 3\alpha)(12+ 5\alpha)},
f_2= \frac{16 i \sqrt{\alpha}}{\sqrt{15}(12 + 5\alpha)},
 f_3=\frac{60 + 16\,i \sqrt{15 \alpha} - 15\alpha}{180 + 75\alpha}
$. We have $f_i\sim o(1)$ for $\alpha\sim 1$, on the other hand
$(f_1,f_2,f_3)\to (0,0,1/3)$ in the limit $\alpha\to 0$.

We insert these solution and
next proceed to diagonalize the neutrino mass matrix; we are
specially interested on the resulting {\em solar} and {\em atmospheric} $\delta m^2$'s, that are:
\begin{eqnarray}
\delta m^2_{sun}&=&\frac{v_5^4\epsilon^4\beta_1^4}{25 M_{\rho_0}^2} h_1(\alpha),\\
\delta m^2_{atm}&=&\frac{v_5^4\epsilon^4\beta_1^4}{25 M_{\rho_0}^2} h_2(\alpha)\,;
\end{eqnarray}
with the functions
$
h_1(\alpha)=(288 \left(5 \alpha ^2-168 \alpha
   +80\right))/(\alpha  (5 \alpha +12)^2),
h_2(\alpha)=(64 \left(15 \alpha ^2-376 \alpha
   +240\right))(\alpha  (5 \alpha +12)^2)
$.
The experimental values \cite{Fogli:2008cx}
$\delta m^2_{sun}\simeq 8\times 10^{-5}\ eV^2$,
$\delta m^2_{atm}\simeq 2.5\times 10^{-3}\ eV^2$
are reproduced if we take:
\begin{eqnarray}
\label{eq:nufit}
M_{\rho_0} \simeq 25\,v_5\,\epsilon^2\,\beta_1^2&,\quad &
\alpha\simeq 1/2.   %%another solution is \quad \mbox{or}\ 32.5: 
\end{eqnarray}
Most interestingly, the hierarchy between the solar and atmospheric
scales only depends on the ratio of the triplet to singlet $\bf 24_T$ fields,
$M_{\rho_3}/M_{\rho_0}$
as:
\begin{eqnarray}
 \frac{ \delta m^2_{sun}}{\delta m^2_{atm}}&=&
(\simeq 3\times 10^{-2}\ (exp))\sim 1-2 \alpha, \\
\frac{M_{\rho_3}}{M_{\rho_0}}&\sim & \frac{1}{2} \left ( 1-
\frac{ \delta m^2_{sun}}{\delta m^2_{atm}} \right ).
\end{eqnarray}
We also ``predict'' the value of the neutrino masses by inserting the fitted values for
$M_\rho$ and $\alpha$ in eq.(\ref{eq:nufit}) in the neutrino mass matrix,
we obtain the following:
\begin{eqnarray}
m_1\simeq m_2&\simeq&0.10\, eV\,,\\
m_2^2-m_1^2&\simeq&8 \times 10^{-5}\, eV^2\,,\\
m_3&=&0.089\, eV\,;
\end{eqnarray}
thus an inverted hierarchy is predicted with an absolute mass scale $\sim 10^{-1}$ eV.

With a quasi degenerated inverted hierarchy the renormalization group
running is especially critical: the neutrino masses can change 
up to a factor $2$, and the mixing angles can receive sizable contributions
\cite{Xing:2006sp}. However the full study of the RGE is beyond the scope of this
paper and it will be treated in full detail in a next publication
\cite{next}.

\section{Conclusions}
We have analyzed all the possible extensions of the recently proposed
minimal renormalizable SUSY $SU(5)$ grand unified model
\cite{Perez:2007iw} with the inclusion of an additional
$A_4$ flavor symmetry.
We have found that there are 5 possible $A_4$ charge assignment cases for the
superfields compatible which $A_4$ invariance.
Among these Cases we found one that is
phenomenologically interesting for both charged fermion and neutrinos masses and mixings,
despite the highly non triviality of the $A_4$ constraints.
In this case all, matter and Higgs, fields (except the ${\bf24_H}$) are triplets under $A_4$,
the field  content and charge assignment are given by:
\begin{center}
\begin{tabular}{||l||c|c|c|c|c|c|c|c||}
\hline\hline
$SU(5)$&${\bf10_T}$ & ${\bf\B5_T}$ & ${\bf24_T}$ & ${\bf\B5_H}$ & ${\bf\B{45}_H}$ & ${\bf5_H}$ & ${\bf45_H}$ & ${\bf24_H}$\\
\hline
\hline
$A_4$&3&3&3&3&3&3&3&1\\
\hline\hline
\end{tabular}
\end{center}

We have studied in detail such Case and
showed how the fermion masses and mixing angles come out.
In particular we arrived to the conclusion that to reproduce the observed
masses and mixing angles the Higgs fields must acquire
their vevs along some particular directions
basically (described in detail in  sec. \ref{sec:scenarioC}) of the form
$\langle {\bf{\B5_H}}\rangle\propto\langle {\bf5_H}\rangle\propto (1,0,0)$,
$\langle {\bf\B{45}_H}\rangle=(1,1,1)$,
and $\langle{\bf{45_H}}\rangle=(1,\epsilon,\epsilon)$, i.e
the $\bf{\B{45}_H}$ vevs breaks down
the $A_4$ symmetry to $Z_3$, while
 the ${\bf 5_H,45_H,\B5_H}$ spontaneous breaking reduce it further down to $Z_2$.

We have obtained that all the experimental charged fermion masses,
quark and lepton mixing angles can be easily fitted.
The absolute scale of  neutrino masses is obtained  as a prediction,
from the experimental TBM mixing neutrino matrix and from the values of
solar and  atmospheric squared mass differences, they are all nearly degenerate
with $m\sim 0.1\,eV$ with an inverted hierarchy.

Let us now compare our model with two other recently proposed SUSY $SU(5)$
models with \cite{Altarelli:2008bg} and without flavor symmetry \cite{Perez:2007iw}.

Let us note first that in the model introduced here,
in contraposition to  \cite {Altarelli:2008bg},
it is necessary  to introduce ad-hoc $U(1)$ and $Z_N$ symmetries to fit
together the discrete and grand  unified symmetries.

On the other hand, in the SUSY ``Adjoint'' $SU(5)$
of \cite{Perez:2007iw}  a full study of the neutrino sector
masses and mixing has not been performed as the one performed here,
so it is not possible to compare  directly both works. It is simply
argued there that is possible  to generate all fermion  masses including
neutrino masses because the Yukawa matrices are arbitrary are the number
of  free parameters is very large.

With respect to that model, which serves of basis of the one presented here,
the inclusion of an extra flavor symmetry of type  $A_4$ implies the following distinctive features:

a) In the D-E sector, down quarks and charged leptons, can be fitted with
 only four parameters, while in a general SUSY $SU(5)$ model we would have
18 parameters. This becomes very much explicit in the combination of matrices:
each of the linear  combination $M_D-M_E^t$, $3 M_D+M_E^t$
depends disjointly  on only two  of the four possible parameters.

b) The up sector contains three parameters which fit well the three masses.
Overall, the $D-E$ and $U$ sectors contains 7 parameters which fits well the
 9 masses and the CKM matrix.
 Moreover, the $D-E,U$ and ``$\nu$'' sectors contain 9 effective parameters which seems
to fit well the 12 masses, the CKM and the PMNS matrices.

c) In both models neutrino masses are coming from a I+III SeeSaw Mechanism. However
here all the neutrinos are massive, while in the other model there is one massless
neutrino. This is due to the form of the low energy mass matrix,
in our model the presence of Higgs $A_4$ triplets implies the
equation  given by Eq.(\ref{eq:nu}).

d) The Tribimaximal neutrino mixing can be easily accommodated.
The absolute scale of  neutrino masses  (they are all nearly degenerate with
$m\sim 0.1\,eV$ with an inverted hierarchy) is obtained
as a prediction,  from the experimental TBM mixing neutrino matrix and
from the values of solar and atmospheric squared mass differences.

So, in conclusion, the introduction of the $A_4$ flavor symmetry seems to be
a ``good'' idea:
\begin{itemize}
\item it restricts very much the number of free parameters, giving simple
 expressions for the Yukawa matrices.
\item The number of Higgs is larger (Higgses are $A_4$ triplets) but their
  vevs are also very much restricted by the $A_4$ symmetry, so the number of
  free parameters coming from this sector does not increase (only 3 parameters).
\item the number of remaining parameters and the texture given to the
 yukawa matrices is big enough to be able to fit the low energy fermion sector
 including the largely different structures of masses and mixing.
\end{itemize}
Two features, at least, of the $A_4$ symmetry  could be tested at low energies in  present and future experiments.
The predicted absolute mass scale and in particular
the mass of the electron neutrino lies well within the sensitivity of near
 future kinematical and double $\beta$ experiments.
The observation of a degenerate neutrino mass spectrum, as the one predicted
 here, could help to differentiate this model with respect other SUSY $SU(5)$
  $A_4$ models, in  particular with \cite{Altarelli:2008bg}.

In summary, we have showed that a simple, TB-Compatible,
 flavor symmetry exclusively based on $A_4$ is compatible with a Grand Unification scenario.
 Moreover, the model considered here can be considered as the simplest, realistic,
  renormalizable supersymmetric grand unified theory of flavor based on the SU(5) gauge
symmetry and $A_4$ flavor symmetry since it has the minimal number of fields and  space-time dimensions.
The details of Higgs vacuum alignment seem to be important for the predictivity of the
 model, and specific to the  discrete flavor symmetry, they will be treated in  full 
 detail in a next publication \cite{next}. In particular it might be interesting  to 
 delucidate which features of this vacuum alignment are specific to the $A_4$ symmetry
and which ones to the mere existence of a discrete  symmetry. 

\vspace{0.4cm}

{\bf Acknowledgments:} We
 acknowledge the partial support provided the Ministerio de Educacion 
 y Ciencia (Goverment of Spain), Fundacion Seneca (Government of the 
 Comunidad Autonoma de Murcia) and the agreement CYCIT-INFN (Spain-Italy)
  through the projects FIS6224-2007, CARM0034-2009 and CYCIT-INFN/12-2008.

\providecommand{\href}[2]{#2}\begingroup\raggedright\endgroup

\end{document}